\documentclass[fleqn,usenatbib]{mnras}

\usepackage{newtxtext,newtxmath}

\usepackage[T1]{fontenc}
\usepackage{ae,aecompl}

\usepackage{graphicx}	
\usepackage{amsmath}	
\usepackage{amssymb}	

\def\f{\frac}

\def\ln{\mathrm{ln}}

\def\cov{\mathbf{C}}

\def\data{\mathbf{d}}

\def\btheta{\boldsymbol{\theta}}

\def\mean{\boldsymbol{\mu}}

\def\transpose{\mathrm{T}}
\def\t{\mathbf{t}}
\def\L{\mathcal{L}}

\title[Scalable likelihood-free inference for cosmology]{Massive optimal data compression and density estimation for scalable, likelihood-free  inference in cosmology}

\author[J. Alsing, B. Wandelt, S. Feeney]{
Justin Alsing,$^{1,2}$\thanks{E-mail: jalsing@flatironinstitute.org}
Benjamin Wandelt$^{1,3,4,5}$
and Stephen Feeney$^{1}$
\\
$^{1}$Center for Computational Astrophysics, Flatiron Institute, 162 5th Ave, New York City, NY 10010, USA\\
$^{2}$Imperial Centre for Inference and Cosmology, Department of Physics, Imperial College London, Blackett Laboratory,\\  Prince Consort
Road, London SW7 2AZ, UK\\ 
$^{3}$Institut d'Astrophysique de Paris (IAP), UMR 7095, CNRS UPMC Universite Paris 6, Sorbonne Universitse,\\ 98bis boulevard Arago, F-75014 Paris, France\\
$^{4}$Institut Lagrange de Paris (ILP), Sorbonne Universitse, 98bis boulevard Arago, F-75014 Paris, France\\
$^{5}$Department of Physics and Astronomy, University of Illinois at Urbana-Champaign, 1002 W Green St, Urbana, IL 61801, USA
}

\date{Accepted XXX. Received YYY; in original form ZZZ}

\pubyear{2017}

\begin{document}
\label{firstpage}
\pagerange{\pageref{firstpage}--\pageref{lastpage}}
\maketitle

\begin{abstract}
Many statistical models in cosmology can be simulated forwards but have intractable likelihood functions. Likelihood-free inference methods allow us to perform Bayesian inference from these models using only forward simulations, free from any likelihood assumptions or approximations. Likelihood-free inference generically involves simulating mock data and comparing to the observed data; this comparison in data-space suffers from the curse of dimensionality and requires compression of the data to a small number of summary statistics to be tractable. In this paper we use massive asymptotically-optimal data compression to reduce the dimensionality of the data-space to just one number per parameter, providing a natural and optimal framework for summary statistic choice for likelihood-free inference. Secondly, we present the first cosmological application of Density Estimation Likelihood-Free Inference (\textsc{delfi}), which learns a parameterized model for joint distribution of data and parameters, yielding both the parameter posterior and the model evidence. This approach is conceptually simple, requires less tuning than traditional Approximate Bayesian Computation approaches to likelihood-free inference and can give high-fidelity posteriors from orders of magnitude fewer forward simulations. As an additional bonus, it enables parameter inference and Bayesian model comparison simultaneously. We demonstrate Density Estimation Likelihood-Free Inference with massive data compression on an analysis of the joint light-curve analysis supernova data, as a simple validation case study. We show that high-fidelity posterior inference is possible for full-scale cosmological data analyses with as few as $\sim 10^4$ simulations, with substantial scope for further improvement, demonstrating the scalability of likelihood-free inference to large and complex cosmological datasets.
\end{abstract}




\section{Introduction}
%
In cosmological data analysis we are often faced with scenarios where we can generate mock data with sophisticated forward simulations, but are unable to write down a tractable likelihood function. For example, physics associated with non-linear structure formation on small scales \citep{Springel2005, Klypin2011}, baryonic feedback \citep{Hellwing2016, Springel2017, Chisari2018}, gravitational and hydrodynamical evolution of the intergalactic medium \citep{Arinyo2015, Bolton2016}, epoch of reionization \citep{Mesinger2016, Kern2017} etc., may be captured (to varying degrees) by simulations, whilst compact and accurate models for the statistical properties of these processes are often elusive. Similarly on the measurement side, complicated noise models, subtle measurement and selection biases etc., can often be simulated but are challenging to incorporate exactly into a likelihood function. 

The standard approach is to build an approximate likelihood that tries to capture as much of the known physics and measurement processes underlying the data as possible, in the hope that the adopted approximations do not lead to biased posterior inferences. Even if the means and variances of inferences are not appreciably biased, assessing tensions between data sets \citep{Marshall2004}, combining inferences and comparing models can be strongly affected by posterior tail probabilities that are unlikely to be accurate when using popular likelihood approximations. With widely reported tensions between key state-of-the art cosmological datasets, most notably weak lensing and cosmic microwave background (CMB) measurements of the amplitude of matter clustering \citep{Planck2015XIII,Joudaki2016,Alsing2016,Hildebrandt2017} and local versus CMB measurements of the Hubble constant \citep{Riess2011,Planck2015XIII,Feeney2017}, it is worthwhile seeking methods that might eliminate likelihood approximations from the chain of scientific reasoning.

Likelihood-free inference methods allow us to perform Bayesian inference using forward simulations only, free from any likelihood assumptions or approximations (see \citealp{Lintusaari2017} for a review). This approach has great appeal for cosmological data analysis, since encoding complex physical processes, instrumental effects, selection biases etc., into a forward simulation is typically much easier than incorporating these effects into a complicated likelihood function and solving the inverse problem.

Likelihood-free methods are emerging as a viable way forward for analyzing complex data-sets in cosmology, with recent applications to inference of the quasar luminosity function \citep{Schafer2012}, galaxy merger rate evolution at early times \citep{Cameron2012}, cosmological parameters from supernova observations \citep{Weyant2013}, galaxy-formation \citep{Robin2014}, weak-lensing peak statistics \citep{Lin2015}, the galaxy-halo connection \citep{Hahn2017}, cosmological redshift distributions \citep{Kacprzak2017}, photometric evolution of galaxies \citep{Carassou2017}, and Lyman-$\alpha$ and -$\beta$ forests \citep{Davies2017}, with public likelihood-free inference codes implementing Approximate Bayesian Computation (\textsc{abc}) facilitating the rise in popularity of these methods \citep{Ishida2015, Akeret2015,Jennings2016}.

In its simplest form, likelihood-free inference with \textsc{abc} involves forward simulating mock data given a set of input parameters drawn from the prior, and then comparing the simulated data to the observed data, accepting parameters when the simulated data is close (by some distance-metric) to the observed data. This comparison in data-space suffers from the curse of dimensionality, scaling exponentially with the size of the data set; for large data sets such a comparison is completely impractical and it is essential to compress the data down to a small number of summary statistics. The need for data compression is common amongst even sophisticated likelihood-free inference methods. Data compression schemes should be carefully designed to reduce the data to the smallest set of summaries possible, whilst retaining as much information about the parameters of interest as possible (see \citealp{Blum2013} for a review).

Once a data compression scheme has been prescribed, the second hurdle for achieving scalable likelihood-free inference is choosing how to propose parameters and run forward simulations in the most efficient way, minimizing the number of simulations required to obtain high-fidelity posterior inferences. This is of particular importance for applications in cosmology, where forward simulations are often extremely computationally expensive; even with very aggressive data compression, \textsc{abc} methods typically require an unfeasibly large number of simulations for many cosmological applications.

This paper tackles the two key hurdles for scalable likelihood-free inference: (1) how do we compress large cosmological datasets down to a small number of summaries whilst retaining as much information about the cosmological parameters as possible, and (2) how do we perform likelihood-free inference using a feasible number of forward simulations. We propose a general two-step data compression scheme, first compressing the full dataset $\mathbf{D}\in\mathbb{R}^N$ down to $\mathbf{d}\in\mathbb{R}^M$ well-chosen heuristic summary statistics as is standard practice in cosmological data analysis (e.g., compressing maps down to power spectra, supernova lightcurves and spectra down to estimated distance moduli and redshifts), and secondly asymptotically-optimally\footnote{For a discussion of precisely what is meant by optimal in this context, see \S \ref{sec:score_compression}.} compressing the $M$ summaries down to just $n$ numbers $\t\in\mathbb{R}^n$ -- one for each parameter of interest -- while preserving the Fisher information content of the data, following \citet{Alsing2017}, \citet{Heavens2000a} and \citet{Tegmark1997}.

With the two-step compression scheme defined, we then introduce Density Estimation Likelihood-Free Inference (\textsc{delfi}; \citealp{Bonassi2011,Fan2013,Papamakarios2016}), which learns a parameterized model for the joint density of the parameters and compressed statistics $P(\btheta, \t)$, from which we can extract the posterior density by simply evaluating the joint density at the observed data $\t_o$, ie., $P(\btheta | \t_o)\propto P(\btheta, \t=\t_o)$. We will show that high-fidelity posterior inference can be achieved with orders of magnitude fewer forward simulations than, for example, available implementations of  Population Monte Carlo ABC (\textsc{pmc-abc}), making likelihood-free inference feasible for full-scale cosmological data analyses where simulations are expensive. As a case study, we will demonstrate \textsc{delfi} with massive data compression on an analysis of the Joint Lightcurve Analysis (JLA) supernova dataset \citep{Betoule2014}.

The structure of this paper is as follows: in \S \ref{sec:compression} we discuss massive asymptotically-optimal data compression for application to likelihood-free inference methods. In \S \ref{sec:lfi} we introduce likelihood-free inference methods, discussing \textsc{abc} and introducing Density Estimation Likelihood-Free Inference or \textsc{delfi} \citep{Bonassi2011,Fan2013,Papamakarios2016} as a scalable alternative. In \S \ref{sec:jla} we validate the \textsc{delfi} method with asymptotically-optimal data compression on a simple analysis of the JLA supernova dataset, and compare to \textsc{pmc-abc}. We conclude in \S \ref{sec:conclusions}.
\section{Massive asymptotically-optimal data compression}
\label{sec:compression}
In this section we describe the two-step data compression, firstly from $N$ data to a set $M$ of well-chosen summary statistics, and then compressing the $M$ summaries down to just $n$ numbers, where $n$ is equal to the number of parameters to be inferred. 

The first step is already common practice in cosmological data analysis; for example, data from cosmological surveys are typically compressed to maps and then estimated $n$-point statistics (power spectra or correlation functions, bispectra etc) or other summary statistics. The second step, compressing down to just one number per parameter whilst retaining as much (Fisher) information about the parameters as possible, has been considered by \citet{Tegmark1997}, \citet{Heavens2000a} and \citet{Alsing2017}; we follow the most general of these studies here \citep{Alsing2017}. These Fisher-information preserving data compression ideas are already widely used in astronomy and cosmology, with applications spanning determining galaxy star formation histories \citep{Reichardt2001, Heavens2004, Panter2007}, CMB data analysis \citep{Gupta2002,Zablocki2016}, gravitational waves \citep{Graff2011}, transient detection \citep{Protopapas2005}, fast covariance matrix estimation \citep{Heavens2017}, galaxy power spectrum and bispectrum analyses \citep{Gualdi2017} and optimal power spectrum estimation \citep{Tegmark1997, Bond1998, Bond2000}.
\subsection{Two-step data compression}
\label{sec:summary_stats}
The two-step compression proceeds as follows:\\

(I) Compress the full dataset $\mathbf{D}\in \mathbb{R}^N$ into a set of summary statistics $\data\in\mathbb{R}^M$, with the aim of retaining as much information as possible about the parameters of interest:
\begin{align}
\mathbf{D} \rightarrow \data(\mathbf{D}) = \{\mathrm{list\;of\;summary\;statistics}\}
\end{align}

(II) Compress the vector of $M$ summary statistics $\data$ into a vector of $n$ numbers $\t\in\mathbb{R}^n$, as follows: assume an approximate form for the log-likelihood function $\mathcal{L}$, and define the compressed statistics $\t$ to be the score-function -- the gradient of the log-likelihood -- evaluated at some fiducial parameter set $\boldsymbol{\theta}_*$:
\begin{align}
\label{linear_compression}
\data \rightarrow \t = \nabla_{\btheta}\L_*.
\end{align}
In the case where we assume a Gaussian likelihood for the summary statistics, $2\L = -(\data-\mean)^\transpose\cov^{-1}(\data-\mean) - \ln|\cov|$ with the mean and covariance depending on the parameters, the compressed statistics $\t$ are given by \citep{Alsing2017}:
\begin{align}
\label{gauss_dtwidle}
\t = \nabla_{\btheta}\mean_*^\transpose\cov^{-1}_*(\data - \mean_*) + \f{1}{2}(\data - \mean_*)^\transpose \cov^{-1}_*\nabla_{\btheta} \cov_*\,\cov_*^{-1}(\data - \mean_*),
\end{align}
where $\mean_*\equiv\mathrm{E}_{\btheta_*}\left[\data\right]$ and $\cov_* = \mathrm{E}_{\btheta_*}\left[(\data-\mean)(\data-\mean)^\transpose\right]$ are the mean and covariance of the summary statistics $\data$ (evaluated at the fiducial parameter values), which can be estimated from forward simulations.

Let us consider the practical considerations and limitations of steps (I) and (II) in turn.
\subsection{Step I: Initial compression to summary statistics}
Compression step (I) retains as much information as can be captured by the carefully chosen summary statistics $\data$. In special cases where \emph{sufficient} summary statistics can be found, this step will be lossless, but in general step (1) will be lossy. Nevertheless, it is standard practice to reduce large cosmological datasets to sets of summary statistics in the hope of retaining as much information as possible whilst making the subsequent inference task tractable. A substantial body of literature exists on summary statistic choice for cosmological data analysis. In the context of likelihood-free inference, one may include in $\data$ a list of as many relevant summary statistics for the problem as is feasible in the hope of capturing as much information as possible, without needing to be able to write down a joint likelihood-function for the summaries.

Importantly, the only requirement for likelihood-free inference is that realizations of the summary statistics can be generated by forward simulation given a set of input parameters; in contrast to likelihood-based analyses, we do not require a predictive model for the expected summary statistics $\boldsymbol\mu(\btheta) = \mathrm{E}_{\btheta}\left[\data\right]$. For cosmological data analysis this has great appeal, since many problems have summary statistics that are expected to contain a wealth of information but that we may not have a reliable predictive model for. Examples in the context of large-scale structure analyses include the galaxy power spectrum and bispectrum in redshift-space and on small scales, the weak lensing power spectrum on small angular scales, cosmic void statistics, the flux power spectrum of the Lyman-$\alpha$ forest, and many others.

Whilst step (I) typically results in an enormous reduction in the size of the data-space, for cosmological applications the number of summary statistics $M$ is still often $\sim10^2$ or larger. For example, we may have compressed a vast number of time-ordered data points from a cosmic microwave background survey down to a few hundred or thousand estimated power spectrum modes, or measured supernova lightcurves and spectra down to an estimated apparent magnitude, redshift, color and stretch parameter for each of the sources. Hence the space of $M$ summaries is still typically much too large for practical data-space comparisons and likelihood-free inference; further compression of these summaries down to a small number of compressed statistics is still required.
\subsection{Step II: Asymptotically optimal compression to the score-function of an approximate likelihood}
\label{sec:score_compression}
Once an appropriate set of summary statistics has been chosen, the massive compression in step (II) proceeds by assuming an approximate form for the likelihood and compressing to the score-function -- the gradient of the log-likelihood evaluated at some fiducial parameter set $\btheta_*$ (Eq. \ref{linear_compression}). This compression results in just $n$ numbers -- one per parameter of interest -- that are optimal in the sense that they preserve the Fisher information, to the extent that the assumed form for the (unknown or intractable) likelihood is a good approximation to the true likelihood function and the fiducial expansion point is close to the maximum-likelihood (this can be iterated if necessary; see \citealp{Alsing2017} for details). A detailed discussion of optimality in this context is given at the end of this section.

Crucially, the assumed approximate likelihood function is used for the sole purpose of performing the data compression; once the compression is done, all likelihood assumptions are dropped and the subsequent inference is genuinely likelihood-free. Better likelihood approximations for the compression step will lead to more optimally compressed statistics, but there is no sense in which these choices will bias the final parameter inferences.

For many applications, a Gaussian likelihood may be a reasonable first approximation, but not accurate enough in detail to be used for likelihood-based inferences. In these situations a Gaussian likelihood may be assumed for the data compression, leading to compressed statistics given by Eq. \eqref{gauss_dtwidle}. Computing the compression in Eq. \eqref{gauss_dtwidle} requires an estimate of the mean, covariance matrix and their derivatives at some fiducial parameter set $\btheta_*$. These can all be obtained using forward simulations only, as is already common practice for mean and covariance estimation for conventional likelihood-based analyses. Derivatives can be estimated quickly by performing simulations with matched random seeds but perturbed parameter values.

For likelihood-based analyses assuming Gaussian likelihoods, poorly estimated covariance matrices will in general lead to biased parameter inferences and great care needs to be taken to ensure these are determined precisely, including any parameter dependencies, and any covariance-matrix uncertainties should be formally marginalized over \citep{Sellentin2015}. In contrast, for likelihood-free analyses, if the covariance matrix used for the compression in Eq. \eqref{gauss_dtwidle} is approximate, the worst outcome is that the resulting compression will be sub-optimal; crucially, there is no sense in which poorly estimated covariance matrices can bias the final parameter inferences and parameter dependent covariance matrices are not required. The same principle applies to the mean and derivatives appearing in Eq. \eqref{gauss_dtwidle}; approximations are always safe, only leading to sub-optimality. This means that when fast approximate models for the covariances and means are available, they can be used safely for rapid compression, reducing the total number of forward simulations required (at the cost of some optimality).
\subsection*{Asymptotic optimality}
Compression to the score function of a given likelihood promises to be optimal in the sense that it preserves the Fisher information content of the data, to the extent that the likelihood assumed for the compression is a good approximation to the true likelihood and the gradient (score) can be evaluated close to the maximum likelihood. 

In typical likelihood-free inference applications, the form of the likelihood function under the model is either not known or not computationally tractable, and using an approximate likelihood function for the compression will be lossy. In these cases, if the compression is performed under a ``best guess'' for the true likelihood then there is still a sense in which the compression is ``optimal"; it preserves as much Fisher information as possible under the level of knowledge and resources available for making a good likelihood approximation for the compression. In other words, you can only do as well as your likelihood-ignorance allows. 

Even when the compression is performed under the exact likelihood, compression to the score only promises to preserve the Fisher information content of the data. Whilst this is a clearly-stated definition of optimality, in cases where the likelihood is a highly non-Gaussian or multi-modal function of the parameters the Fisher information is not guaranteed to be a good measure of the information content of the data and there may be more effective compression schemes. This is rarely the case for cosmological applications. Nevertheless, in the asymptotic limit where the likelihood becomes Gaussian with (expected) curvature specified by the Fisher information matrix, compression to the score exactly preserves the (expected) uncertainties on the inferred parameters. In this sense, compression to the score can be said to be asymptotically-optimal.

We note that a new approach to data compression is emerging that does not require the compression to be performed under an approximate likelihood function; \citet{Charnock2018} develop information maximizing neural networks trained on forward simulations that can learn optimal compression schemes without specifying a likelihood function.
\section{Likelihood-free inference}
\label{sec:lfi}
In this section we discuss likelihood-free inference. In \S \ref{sec:abc} we discuss Approximate Bayesian Computation (\textsc{abc}) methods, highlighting some of their limitations in the context of cosmological data analysis. In \S \ref{sec:gmm_lfi} we present Density Estimation Likelihood-Free Inference (\textsc{delfi}), which overcomes many of the key shortcomings of \textsc{abc} methods.
\subsection{Approximate Bayesian Computation (ABC)}
\label{sec:abc}
In its simplest incarnation, rejection \textsc{abc} works as follows \citep{Rubin1984}: 
\begin{enumerate}
\item Draw parameters from the prior $\btheta \leftarrow P(\btheta)$;
\item Simulate mock data $\data \leftarrow P(\data | \btheta)$;
\item If distance between observed and mock data is smaller than some threshold, $\rho(\data, \data_o) < \epsilon$, accept, else reject;
\item Repeat until desired number of samples are obtained. 
\end{enumerate}
In the limit where $\epsilon\rightarrow 0$, the accepted samples are drawn from the true posterior, whilst for any non-zero $\epsilon$, the samples drawn are from an approximate posterior that is by construction broader than the true posterior density. The distance metric $\rho$ for comparing simulated and observed data needs to be specified (with many options existing, \citealp{Mckinley2009}), as does the distance threshold $\epsilon$. Since the acceptance rate becomes vanishingly small as $\epsilon\rightarrow 0$, \textsc{abc} posteriors are always broader than the true posterior, but are unbiased; provided one can make $\epsilon$ small enough, good posterior approximations can be recovered.

Proposing parameters from the prior in rejection \textsc{abc} is typically inefficient when the posterior density occupies a small portion of the total prior volume (see eg., \citealp{Toni2009, ToniStumpf2009}). In this case, drawing parameters from a proposal distribution that preferentially samples the relevant portion of parameter space (followed by importance re-weighting) leads to more efficient \textsc{abc} sampling. Population Monte Carlo (\textsc{pmc}) and Sequential Monte Carlo (\textsc{smc}) \textsc{abc} methods \citep{DelMoral2006, Sisson2007, Beaumont2009, Toni2009, Bonassi2015} are popular advancements on rejection \textsc{abc} that adaptively learn a more intelligent proposal distribution, whilst at the same time implementing a ``cooling" scheme for $\epsilon$, gradually lowering the distance threshold as the proposal distribution becomes more optimized (see \citealp{Ishida2015,Akeret2015,Jennings2016} for applications in the astronomy literature).

\textsc{abc} methods have been applied successfully to a number of problems in cosmological data analysis \citep{Schafer2012,Cameron2012,Weyant2013,Robin2014,Lin2015,Hahn2017,Kacprzak2017,Davies2017}. However, even sophisticated \textsc{abc} algorithms suffer from vanishingly small acceptance rates as $\epsilon \rightarrow 0$ by construction, scaling poorly with the number of parameters of interest, so for high-fidelity posterior inference this usually means running very large numbers of forward simulations. For many applications in cosmology where simulations are expensive, this is impractical.

In the next section we describe a totally different approach to likelihood-free inference that is ``$\epsilon$-free", circumventing the need to do direct comparisons in data-space and ultimately making much more efficient use of forward simulations.
\subsection{Density Estimation Likelihood-Free Inference (DELFI)}
\label{sec:gmm_lfi}
Density Estimation Likelihood-Free Inference works by learning a parameterized model for the joint density $P(\btheta, \data)$, from a set of samples drawn from that density \citep{Bonassi2011,Fan2013,Papamakarios2016}. In its simplest form, we start by generating a set of samples $\{\btheta, \data\}$ from $P(\btheta, \data)$ by drawing parameters from the prior and forward simulating mock data:
\begin{align}
&\btheta \leftarrow P(\btheta) \nonumber \\
&\data \leftarrow P(\data | \btheta).
\end{align}
We then write down a model for the joint density $P(\btheta, \data ; \boldsymbol{\eta})$, parameterized by $\boldsymbol{\eta}$, and fit this model to the samples $\{\btheta, \data\}$. The estimated\footnote{Recall that widely used \textsc{mcmc} methods also produce \emph{estimates} of the posterior density (and its properties) and/or the model evidence, from a set of posterior samples.} posterior density and Bayesian evidence can then be easily extracted from the fit to the joint density as follows:
\begin{align}
\label{slice}
&\hat{P}(\btheta | \data_o) \propto P(\btheta, \data = \data_o;\boldsymbol{\eta}) \nonumber \\
&\hat{P}(\data_o) = \int P(\btheta, \data = \data_o;\boldsymbol{\eta})\,d\btheta,
\end{align}
ie., taking a slice through the joint distribution evaluated at the observed data $\data = \data_o$, and subsequently integrating over the parameters for the Bayesian evidence. For many practical choices of parameterized models for the joint density, eg., Gaussian mixture models (see below), the evidence integral in Eq. \eqref{slice} is analytically tractable. This means that the evidence comes for free, and if the parameterized model for the joint density is fit to the samples in a principled way, the uncertainties on the fit parameters can be propagated through to a principled uncertainty on the estimated Bayesian evidence.
 
In contrast to \textsc{abc}, \textsc{delfi} uses all of the available forward simulations $\{\btheta, \data\}$ to inform the inference of the joint density $P(\btheta, \data)$, and hence the posterior density and evidence estimation. In practice, this means that far fewer forward simulations may be needed to obtain high-fidelity posterior inferences (compared to \textsc{abc} that has a vanishingly small acceptance rate as $\epsilon\rightarrow 0$), as demonstrated by \citet{Papamakarios2016}.
\subsection*{Gaussian mixture density estimation}
In this work we parameterize the joint density with a Gaussian mixture model (\textsc{gmm}),
\begin{align}
\label{gmm_model}
P(\btheta, \data;\boldsymbol{\eta}) = \sum_{i=1}^K w_i\mathcal{N}(\boldsymbol{\mu}_i, \cov_i),
\end{align}
where $\mathcal{N}(\cdot)$ denotes the Gaussian density, and the mixture model is parameterized by the weights, means and covariances $\boldsymbol{\eta}=\{\{w\}, \{\boldsymbol{\mu}\}, \{\cov\}\}$ of each of the $K$ components, with $\sum w = 1$. \textsc{gmm}s are capable of representing any probability density arbitrarily accurately, provided the number of components $K$ is sufficiently large, and are straightforward to fit to data using expectation-maximization or other methods (see eg., \citealp{Bishop2006}). They also have the appeal that the evidence integrals appearing in Eq. \eqref{slice} are analytically tractable, so the Bayesian evidence comes for free:
\begin{align}
\label{gmm_evidence}
\hat{P}(\data_o) = \int P(\btheta, \data = \data_o;\boldsymbol{\eta})\,d\btheta=\sum_{i=1}^K w_i\mathcal{N}(\boldsymbol{\mu}_{\data\,i}, \cov_{\data\data\,i}),
\end{align}
where $\boldsymbol{\mu}_{\data\,i}$ and $\cov_{\data\data\,i}$ are the component means and covariances corresponding to the data dimensions in the joint density.
\subsection*{Data compression}
The need for data compression for \textsc{delfi} is still clear: the joint density of the data and parameters has dimensionality $N+n$, which presents an intimidating density estimation task for even modestly large datasets. However, implementing the two-step compression scheme described in \S \ref{sec:compression} means we only have to estimate the joint-density of the parameters and compressed statistics $P(\t, \btheta)$, whose dimensionality is just $2n$. For many cosmological applications, the number of parameters of interest is typically $n\lesssim 10$.

When using \textsc{delfi} with a data compression scheme, samples $\{\btheta, \t\}$ are generated from $P(\btheta, \t)$ as before by drawing parameters from the prior and forward simulating mock data, with the addition of the subsequent compression step:
\begin{align}
&\btheta \leftarrow P(\btheta), \nonumber \\
&\data \leftarrow P(\data | \btheta), \nonumber \\
&\t = \t(\data).
\end{align}
These samples $\{\btheta, \t\}$ are then fit with a mixture density model in the usual way as described above. Note that when the Bayesian evidence is estimated from the joint density $P(\btheta, \t)$, this will be the evidence for the compressed statistics $P(\t_o)$ and not for the original data vector $P(\data_o)$; whilst these are not numerically equivalent, the evidence under compressed statistics can still be readily used for model comparison purposes provided alternative models are compared under the same set of compressed summaries.

Importantly, the complexity of the inference problem stays as a $2n$-dimensional density estimation task irrespective of the size of the dataset (or the number of first-level summaries used), once the compression scheme has been prescribed. Therefore, the inference step scales easily to large datasets.
\subsection*{Implementation}
We run forward simulations to generate a set of samples $\{\btheta, \t\}$ and fit a \textsc{gmm} using \textsc{pygmmis} \citep{Melchior2016}, which uses expectation-maximization whilst properly taking into account any hard prior boundaries. 

Gaussian mixture density-estimation with a large number of components can fall foul to over-fitting. One simple way to mitigate over-fitting is to set a minimum threshold for the diagonals of the mixture component covariances (we adopt this approach). For a more sophisticated implementation that avoids over-fitting without having to specify thresholds by hand, see \citet{Papamakarios2016}.

Note that when there are hard prior boundaries, the evidence integral in Eq. \ref{gmm_evidence} is no longer analytically tractable. In these cases, one can estimate the evidence as follows: fit a \textsc{gmm} to the samples of $\{\t\}$ alone, ignoring $\{\btheta\}$ (this effectively pre-marginalizes over $\btheta$). Then the evidence can be estimated by simply evaluating at the observed data, ie., $\hat{P}(\t=\t_o)$.
\subsection*{Sophistications}
\citet{Papamakarios2016} developed a sophisticated implementation of \textsc{delfi} with two key advancements on the vanilla set-up described above. Firstly, they parameterize the joint distribution with a mixture density network (\textsc{mdn}) -- a neural network parameterization of a Gaussian mixture model -- which is fit to the samples using stochastic variational inference (\textsc{svi}; see \citealp{Bishop2006} for a review of \textsc{mdn} and \citealp{Hoffman2013} for \textsc{svi} methods). Secondly, rather than drawing samples from the prior, they adaptively learn a proposal distribution that preferentially samples regions of high posterior density, and subsequently importance re-weight the samples (in the same spirit as population monte-carlo \textsc{abc} methods). They find that this set-up is highly resistant to over-fitting even for small numbers of samples, enabling the number of forward simulations to be reduced further. We leave implementation of these sophistications to future work. 
\subsection*{Scaling with number of parameters and dealing with nuisances}
With the compression scheme employed, the inference task is reduced to learning a $2n$ dimensional density from a set of forward simulations, irrespective of the size of the dataset. The complexity of the inference problem will increase with the number of parameters $n$.

For typical cosmological applications, the number of parameters of interest $\btheta$ (ie., the cosmological parameters) will be $\lesssim 10$. However, in many situations there will be additionally a number of nuisance parameters $\boldsymbol{\xi}$, capturing observational and astrophysical systematics, selection effects etc., which need inferring simultaneously and subsequently marginalizing over. If there are $n$ parameters of interest and $m$ nuisance parameters, \textsc{delfi} involves learning a $2(n+m)$ dimensional probability density over the parameters, nuisances and their respective compressed summaries, $P(\btheta, \boldsymbol{\xi}, \t_\theta, \t_\xi)$. However, if the goal is the posterior marginalized over the nuisance parameters, it may be possible to keep the complexity of the inference task as a $2n$-dimensional density estimation problem. This can be achieved by choosing compressed summaries $\t_\theta\in\mathbb{R}^n$ that contain as much information as possible about the parameters of interest, whilst being as insensitive as possible to the nuisance parameters. Then, draw samples $\{\btheta, \boldsymbol{\xi}, \t_\theta\}$ by drawing from the prior and forward simulating as usual, but only attempt to fit the density $P(\btheta, \t_\theta)$, ie., a $2n$-dimensional density pre-marginalized over the nuisances.

Data compression for marginalized parameter subsets (under Gaussian likelihoods) is treated in \citet{Zablocki2016}. We leave a more general treatment of optimal compression in the presence of nuisances and learning the nuisance-marginalized posterior density to future work.
\section{Validation Case study: JLA Supernova data analysis}
\label{sec:jla}
\begin{figure*}
\centering
\includegraphics[width = 16.5cm]{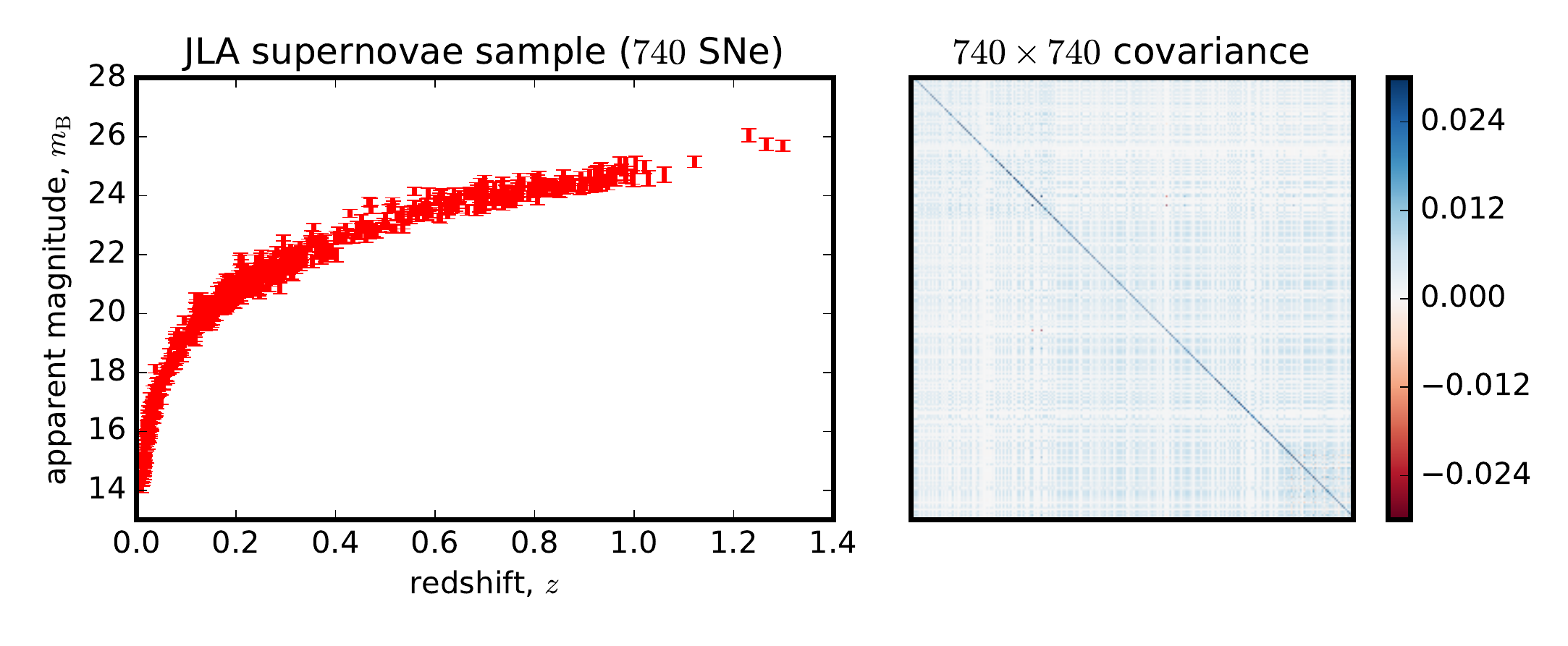}
\caption{Left: Measured apparent magnitudes with their associated uncertainties (from the diagonal of the covariance matrix) for the sample of $740$ supernovae in the JLA sample. Right: covariance matrix of the measured apparent magnitudes, having taken into account redshift and lightcurve calibration uncertainties -- see \citet{Betoule2014} for details of the covariance matrix construction.}
\label{fig:jla_data}
\end{figure*}
To demonstrate the use of Density Estimation Likelihood-Free Inference (\S \ref{sec:lfi}) with massive optimal data compression (\S \ref{sec:compression}), in this section we perform an analysis of the Joint Lightcurve Analysis (JLA) supernova dataset \citep{Betoule2014}. For the purposes of validating the likelihood-free approach, we perform a simple analysis assuming a Gaussian likelihood for the JLA data so that we can compare to an exact likelihood-based analysis, allowing us to demonstrate the fidelity of the likelihood-free posterior inference against a ground-truth\footnote{Note that this simple validation case study is for method-validation purposes only. It is not intended to incorporate new physics or systematics over-and-above the standard JLA analysis of \citet{Betoule2014}.}. We will compare \textsc{delfi} and \textsc{pmc-abc} against a long-run Markov Chain Monte Carlo (\textsc{mcmc}) analysis of the exact (assumed) posterior distribution.

In \S \ref{sec:data}--\ref{sec:jla_likelihood} we describe the JLA data, model and Gaussian likelihood assumptions under which we validate the likelihood-free approach. In \S \ref{sec:implementation} we discuss the implementation of the likelihood-free inference, and in \S \ref{sec:results} we show the results. 
\subsection{JLA supernova data}
\label{sec:data}
We use the JLA sample comprised of observations of $740$ type Ia supernovae, as analyzed in \citet{Betoule2014}. The sample is a compilation of supernova observations from a number of surveys -- see \citet{Betoule2014} and references therein for details.

The full dataset comprises multicolor lightcurves and spectroscopic (or sometimes photometric) observations of each supernova. These lightcurves and spectra are then used to estimate apparent magnitudes $m_\mathrm{B}$ and redshifts $z$, as well as color at maximum-brightness $C$ and stretch $X_1$ parameters characterizing the lightcurves (see eg., \citealp{Tripp1998}). In the data analysis (see \S \ref{sec:jla_likelihood}, also \citealp{Betoule2014}), the data vector will be assumed to be the vector of estimated apparent magnitudes $\data = (\hat{m}_\mathrm{B}^1, \hat{m}_\mathrm{B}^2,\dots,\hat{m}_\mathrm{B}^M)$, where uncertainties in the redshift, color and stretch parameters are propagated through to the covariance matrix of the observed apparent magnitudes. This compression of the multicolor lightcurves and spectra down to a set of estimated apparent magnitudes can be thought of as step (I) of the data compression described in \S \ref{sec:compression}.
\subsection{wCDM and lightcurve calibration model}
As standardizable candles, we assume that the apparent magnitudes of type Ia supernovae depend on the luminosity distance to the source at a given redshift $D^*_\mathrm{L}(z)$ (which is a function of the cosmological model and parameters), a reference absolute magnitude for type Ia supernovae (as a function of host-galaxy mass), and calibration corrections for the light-curve stretch $X_1$ and color at maximum-brightness $C$,
\begin{align}
\label{apparent_mag}
m_\mathrm{B} = 5\mathrm{log}_{10}\left[\f{D^*_\mathrm{L}(z)}{10\mathrm{pc}}\right] + \tilde{M}_\mathrm{B}(M_\mathrm{stellar}; M_\mathrm{B}, \delta M) - \alpha X_1 + \beta C
\end{align}
where $\alpha$ and $\beta$ are calibration parameters for the stretch and color respectively. The absolute magnitude $\tilde{M}_\mathrm{B}$ is assumed to be dependent on the properties of the host galaxy; following \citet{Betoule2014}, we model the dependence of the reference absolute magnitude on the stellar mass of the host as $\tilde{M}_\mathrm{B} = M_\mathrm{B} + \delta M\,\Theta(M_\mathrm{stellar} - 10^{10}M_\odot)$, where $\Theta$ is the Heaviside function.

The cosmological model enters in the luminosity distance-redshift relation. We will assume a flat universe with cold dark matter and dark energy characterized by equation-of-state $p/\rho=w_0$ (hereafter, $w$CDM). In a $w$CDM universe, the luminosity distance is given by,
\begin{align}
D^*_\mathrm{L}(z) = \f{(1+z)c}{100}\int_0^z \f{dz'}{\sqrt{\Omega_\mathrm{m}(1+z')^3 + (1-\Omega_\mathrm{m})(1+z')^{3(w_0+1)}}},
\end{align}
where $\Omega_\mathrm{m}$ is the matter density parameter, $c$ is the speed of light (in vacuum) and $w_0$ is the equation-of-state of dark energy.

The resulting $w$CDM model with color and stretch calibration and host-mass dependent absolute magnitude has six free parameters of interest: $\btheta = (\Omega_\mathrm{m}, w_0, \alpha, \beta, M_\mathrm{B}, \delta M)$.
\subsection{Likelihood}
\label{sec:jla_likelihood}
Following \citet{Betoule2014}, for this validation case we will assume the data $\data = (\hat{m}_\mathrm{B}^1, \hat{m}_\mathrm{B}^2,\dots,\hat{m}_\mathrm{B}^M)$ are Gaussian distributed, 
\begin{align}
\label{jla_likelihood}
\mathcal{L} = -\f{1}{2}(\data - \boldsymbol\mu(\btheta))^T\cov^{-1}(\data - \boldsymbol\mu(\btheta)) - \f{1}{2}\ln |\cov|,
\end{align}
where the mean depends on the parameters and is given by Eq. \eqref{apparent_mag}, and we will assume a fixed covariance matrix\footnote{\citet{Betoule2014} constructed a covariance matrix that depends on $\alpha$ and $\beta$, and also dropped the $|\cov|$ term from the likelihood (Eq. \ref{jla_likelihood}) following \citet{March2011}. However, since $\alpha$ and $\beta$ are very well constrained by the data, the covariance dependence has a small impact on the final parameter inference. For this study, we compute the covariance described in \citet{Betoule2014} but with $\alpha$ and $\beta$ fixed to their maximum likelihood values: $\alpha=0.1257$, $\beta=2.644$. This also avoids issues arising from dropping the $|\cov|$ term from the Gaussian likelihood.}, shown in Fig. \ref{fig:jla_data}, that is assumed to have already accounted for the uncertainties in the color, stretch and redshift of each measured supernova (see \citealp{Betoule2014} for details of the covariance matrix construction).
\subsection{Priors}
We assume broad Gaussian priors on the parameters $\btheta= (\Omega_\mathrm{m}, w_0, \alpha, \beta, M_\mathrm{B}, \delta M)$ with the following mean and covariance:
\begin{align}
&\boldsymbol{\mu}_\mathrm{P} = (0.3,\;  -0.75 ,\; -19.05 ,\;   0.125,\; 2.6  ,  \;-0.05),\nonumber \\
&\cov_\mathrm{P}=
  \left( {\begin{array}{cccccc}
   0.4^2 & -0.24 & 0 & 0 & 0 & 0 \\
   -0.24 & 0.75^2 & 0 & 0 & 0 & 0 \\
   0 & 0 & 0.1^2 & 0 & 0 & 0 \\
   0 & 0 & 0 & 0.025^2 & 0 & 0 \\
   0 & 0 & 0 & 0 & 0.25^2 & 0 \\
   0 & 0 & 0 & 0 & 0 & 0.05^2 \\
\end{array} } \right).
\end{align}
In addition to the Gaussian prior, we impose hard prior boundaries on $\Omega_\mathrm{m}\in[0, 0.6]$ and $w_0\in[-1.5, 0]$. The (truncated) Gaussian prior is much broader than the resulting posterior, having a negligible impact on the the posterior inference relative to (infinite) uniform priors.

The correlations in the Gaussian prior are chosen to roughly follow the correlation structure of the inverse Fisher matrix for the parameters; this allows us to form a broad, weakly informative prior whilst improving the volume ratio of the posterior and prior (ie., giving low prior weight to regions of parameter space that are anticipated to be strongly disfavoured by the likelihood, based on the Fisher matrix). We find this has negligible impact on the parameter inferences whilst improving the performance of the likelihood-free inference methods\footnote{Note that \textsc{pmc abc} and similar \textsc{pmc} approaches to \textsc{delfi} \citep{Papamakarios2016} will be less sensitive to the posterior-prior volume ratio, since they adaptively learn a proposal density rather than blindly proposing parameters from the prior.}.
\subsection{Massive asymptotically-optimal compression}
For step (II) of the data compression, from $N=740$ estimated apparent magnitudes down to $n=6$ numbers (one per parameter, following \S \ref{sec:compression}) we assume a Gaussian likelihood as in Eq. \eqref{jla_likelihood} where only the mean depends on the parameters. In this case, following Eq. \eqref{gauss_dtwidle} \citep{Alsing2017,Heavens2000a} the optimally compressed statistics are given by\footnote{Note that under the assumptions of Gaussian data where the only parameter dependence is in the mean, the compressed statistics are equivalent to the \textsc{moped} linear data compression of \citet{Heavens2000a}.}:
\begin{align}
\t = \nabla_{\btheta}\mean_*^\transpose\cov^{-1}_*(\data - \mean_*) + \cov_{\mathrm{P}}^{-1}(\mean_\mathrm{P} - \btheta_*),
\end{align}
where the mean, its derivative, and the covariance matrix are evaluated at some fiducial point $\btheta_*$ and the second term includes the impact of the Gaussian prior\footnote{This is a minor extension of the derivation in \citet{Alsing2017}, replacing the log-likelihood in $\t=\nabla_{\btheta}\L$ with the sum of the log-likelihood and log-prior, to incorporate the impact of the prior into the optimal compression.}. To choose an optimal fiducial parameter set for the compression, we iterate the parameters using the Fisher scoring method \citep{Alsing2017}:
\begin{align}
\label{fisher_scoring}
\btheta_{k+1} = \btheta_k + \mathbf{F}^{-1}_k\mathbf{t}_k,
\end{align}
where $\mathbf{F}=\nabla\mean^\transpose\cov^{-1}\nabla^\transpose\mean$ is the Fisher information matrix \citep{Tegmark1997}, and $\mathbf{t}_k$ are the compressed statistics computed about the fiducial point $\btheta_k$. We find that Eq. \eqref{fisher_scoring} converges very quickly and gives the following expansion point: $\btheta_*=(0.202, -0.748, -19.04, 0.126, 2.644, -0.0525)$.

The derivatives of the mean with respect to the parameters can be written down analytically for $\alpha$, $\beta$, $M_\mathrm{B}$ and $\delta M$ (see Eq. \ref{apparent_mag}), whilst we use a simple leap-frog approximation for the derivatives with respect to $\Omega_\mathrm{m}$ and $w_0$.
\subsection{Likelihood-free inference implementation}
\label{sec:implementation}
For this validation case we are assuming that the data is Gaussian distributed, with mean given by Eq. \eqref{apparent_mag} and fixed covariance as shown in Fig. \ref{fig:jla_data}. Forward simulating data realizations given parameters is hence as simple as drawing Gaussian random variates, and samples from the joint data-parameter density $\{\btheta, \t\}$ are generated as follows:
\begin{align}
\label{simulation}
&\btheta \leftarrow \mathcal{N}(\boldsymbol{\mu}_\mathrm{P}, \cov_\mathrm{P}), \nonumber \\
&\data = \boldsymbol{\mu}(\btheta) + \mathbf{n},\;\mathrm{where}\;\mathbf{n}\leftarrow \mathcal{N}(\mathbf{0}, \cov_*) \nonumber \\
&\t = \nabla_{\btheta}\mean_*^\transpose\cov^{-1}_*(\data - \mean_*)
\end{align}
We generated a set of $20,000$ samples $\{\btheta, \t\}$ from the joint distribution and fit them with Gaussian mixture models. The \textsc{gmm} fits are performed using expectation-maximization (implemented using \textsc{pygmmis}; \citealp{Melchior2016}) as described in \S \ref{sec:gmm_lfi}, with $K=1$ through to $K=18$ mixture components. The mixture component covariances are regularized by setting a minimum threshold value of $10^{-6}$ for the diagonal values, to avoid over-fitting. We assess convergence with respect to the number of mixture components by looking at the total log-likelihood of the samples under the Gaussian mixture model, as a function of the number of mixture components (see Fig. \ref{fig:convergence}, discussion in \S \ref{sec:results}). Convergence with respect to the number of samples fed to the \textsc{gmm} is assessed by looking for convergence in the recovered posterior means and covariances (see Fig. \ref{fig:posterior_convergence}, discussion in \S \ref{sec:results}).

Since the prior in this case has hard boundaries, the Bayesian evidence integral from the \textsc{gmm} (Eq. \ref{gmm_evidence}) is no longer analytically tractable. We estimate the evidence by fitting a \textsc{gmm} directly to the samples $\{\t\}$ to obtain an estimate of the density $P(\t)$, and then evaluating at the observed data to estimate the evidence $P(\t_o)$. We use the same \textsc{gmm} set-up (ie., number of components) for the evidence as was used for the final posterior inference (discussion in \S \ref{sec:results}).
\subsection{Results}
\label{sec:results}
\begin{figure}
\includegraphics[width = 8.5cm]{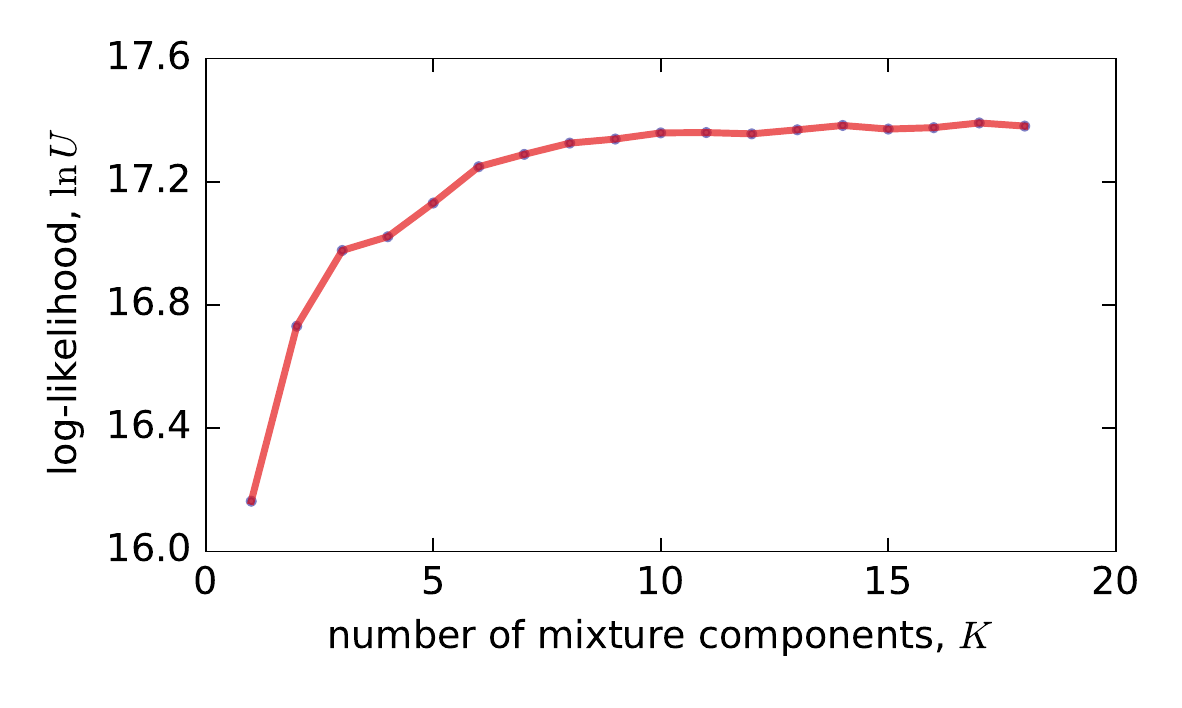}
\caption{Log-likelihood of the samples $\{\t, \btheta\}$ under the \textsc{gmm} fits to the joint density $P(\t, \btheta)$, as a function of the number of Gaussian mixture components $K$. The log-likelihood converges with respect to the number of mixture components, where regularization of the mixture component covariances protects against over-fitting (see \S \ref{sec:implementation}).}
\label{fig:convergence}
\end{figure}
\begin{figure*}
\centering
\includegraphics[width = 17.5cm]{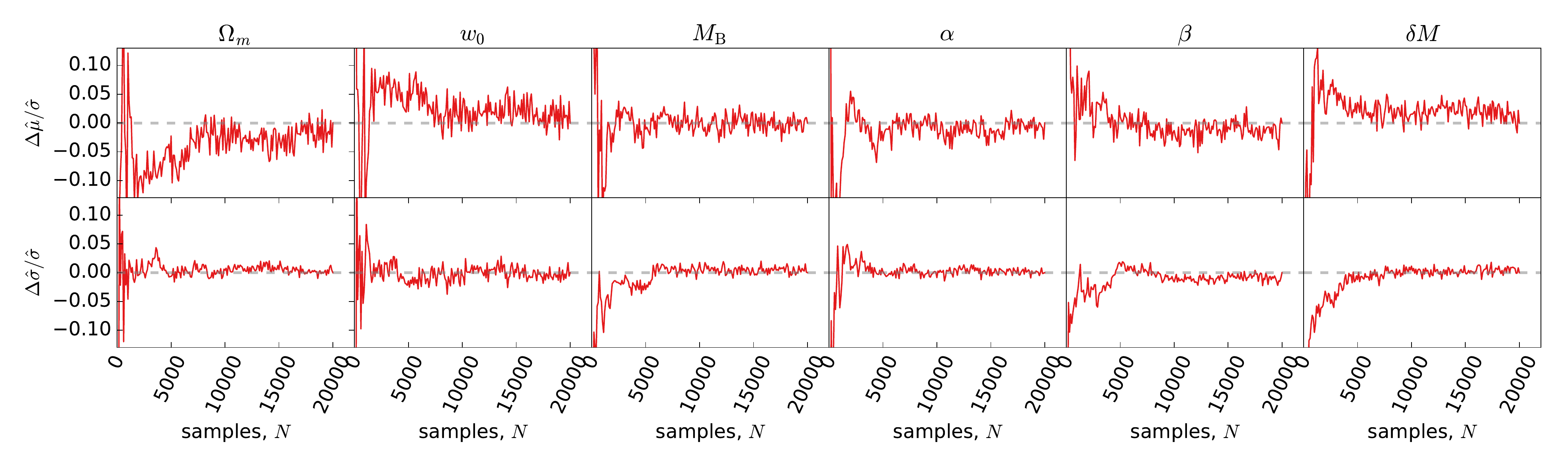}
\caption{Top row: Convergence in the estimated posterior mean as a function of the number of forward simulations fed to the Gaussian mixture model fit to the joint density $\hat{P}(\btheta, \t)$. The panels show $\Delta\hat\mu/\hat\sigma = (\hat\mu_N - \hat\mu_{N=20,000})/\hat\sigma_{N=20,000}$, where $\hat\mu_N$ and $\hat\sigma_N$ are the estimated posterior mean and standard deviation from a Gaussian mixture fit to $N$ forward simulated samples $\{\t, \btheta\}$. The posterior mean converges after a few thousand simulations, with some residual scatter of $\lesssim 0.05\sigma$ for each parameter. Bottom row: Similarly, convergence of the estimated posterior standard deviation for each parameter as a function of the number of simulations fed to the \textsc{gmm}. The standard deviations also converge after a few thousand forward simulations, with some residual scatter at the level of a few percent. Much of the residual scatter in the posterior means and standard deviations is due to small residual \textsc{gmm} fitting uncertainties.}
\label{fig:posterior_convergence}
\end{figure*}
To assess convergence with respect to the number of mixture components, Fig. \ref{fig:convergence} shows the log-likelihood of the samples under the \textsc{gmm} model fits to the joint density, as a function of the number of mixture components $K$ (using the full $20,000$ samples; see below). The log-likelihood of the samples clearly converges with the number of components, reaching a point where adding more components does not improve the fit further. The regularization of the mixture-component covariances has ostensibly protected against over-fitting successfully (cf., \S \ref{sec:implementation}). We use the $K=12$ component \textsc{gmm} model moving forward, which appears to be well into the regime where the log-likelihood of the fit has converged.

Fig. \ref{fig:posterior_convergence} shows convergence of the posterior means and standard deviations (for each parameter) as a function of the number of samples $N$ that are fed to the \textsc{gmm} model fit to the joint density. The posterior means and standard deviations converge after $N\approx 8000$, with some residual scatter of $\lesssim 0.05\sigma$ for each parameter in the means, and a few percent in the standard deviations. This gives confidence that a reasonable posterior approximation may be obtained from just a few thousand forward simulations. However, if we are interested in high-fidelity posterior inference capturing the detailed shape of the posterior, a most sophisticated convergence diagnostic may be required and we leave this to future work. We proceed using the Gaussian mixture model fit to $20,000$ forward simulations to ensure excellent convergence, but note that in practice perfectly adequate posterior inferences may be made with far fewer simulations.
\begin{figure*}
\centering
\includegraphics[width = 17.5cm]{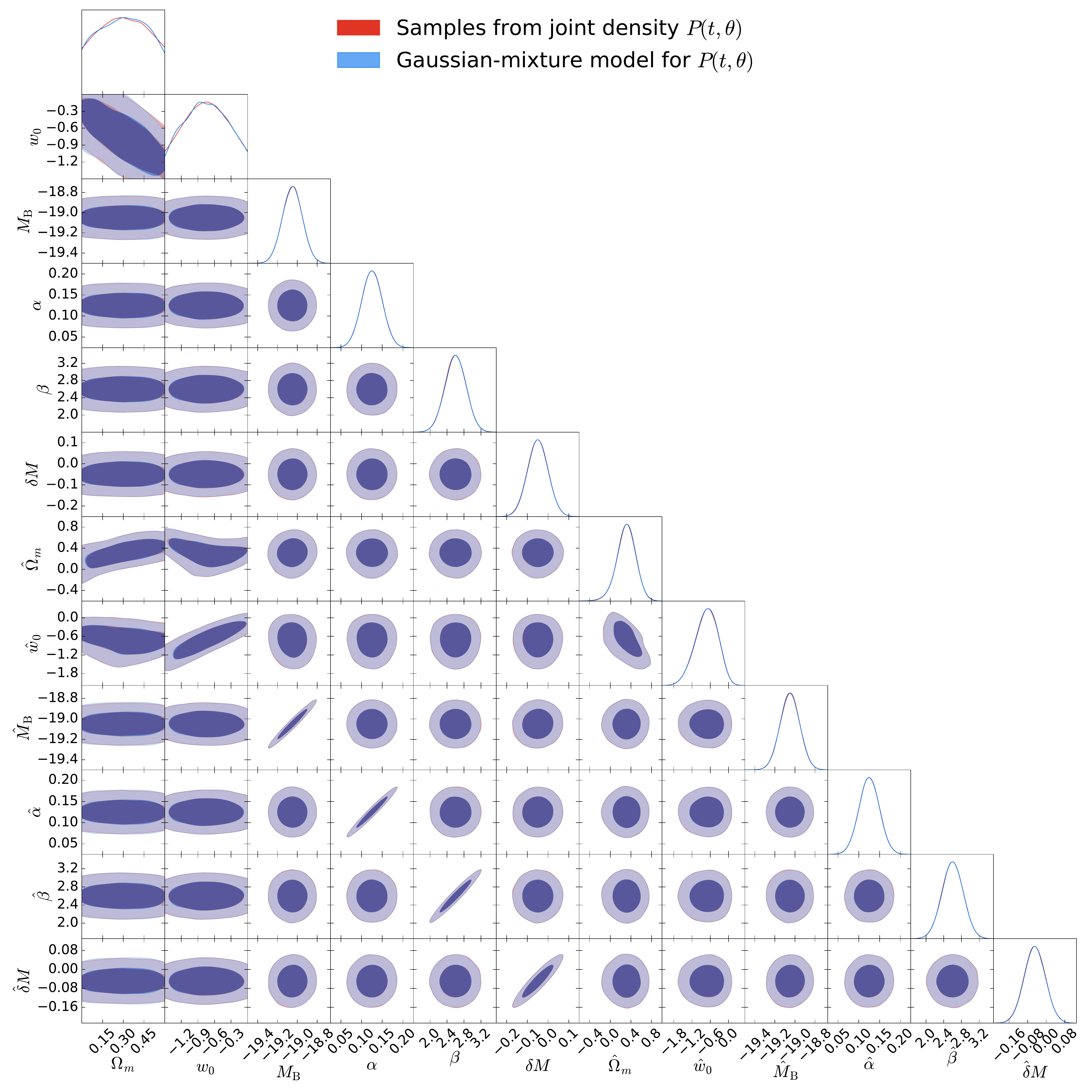}
\caption{Samples from the joint density $P(\t, \btheta)$ (red) against the \textsc{gmm} fit to the samples (blue) for a $K=12$ component mixture model. The \textsc{gmm} model fit provides an excellent representation of the joint density $P(\t, \btheta)$.}
\label{fig:gmm_fit}
\end{figure*}
\begin{figure*}
\centering
 \includegraphics[width = 17.5cm]{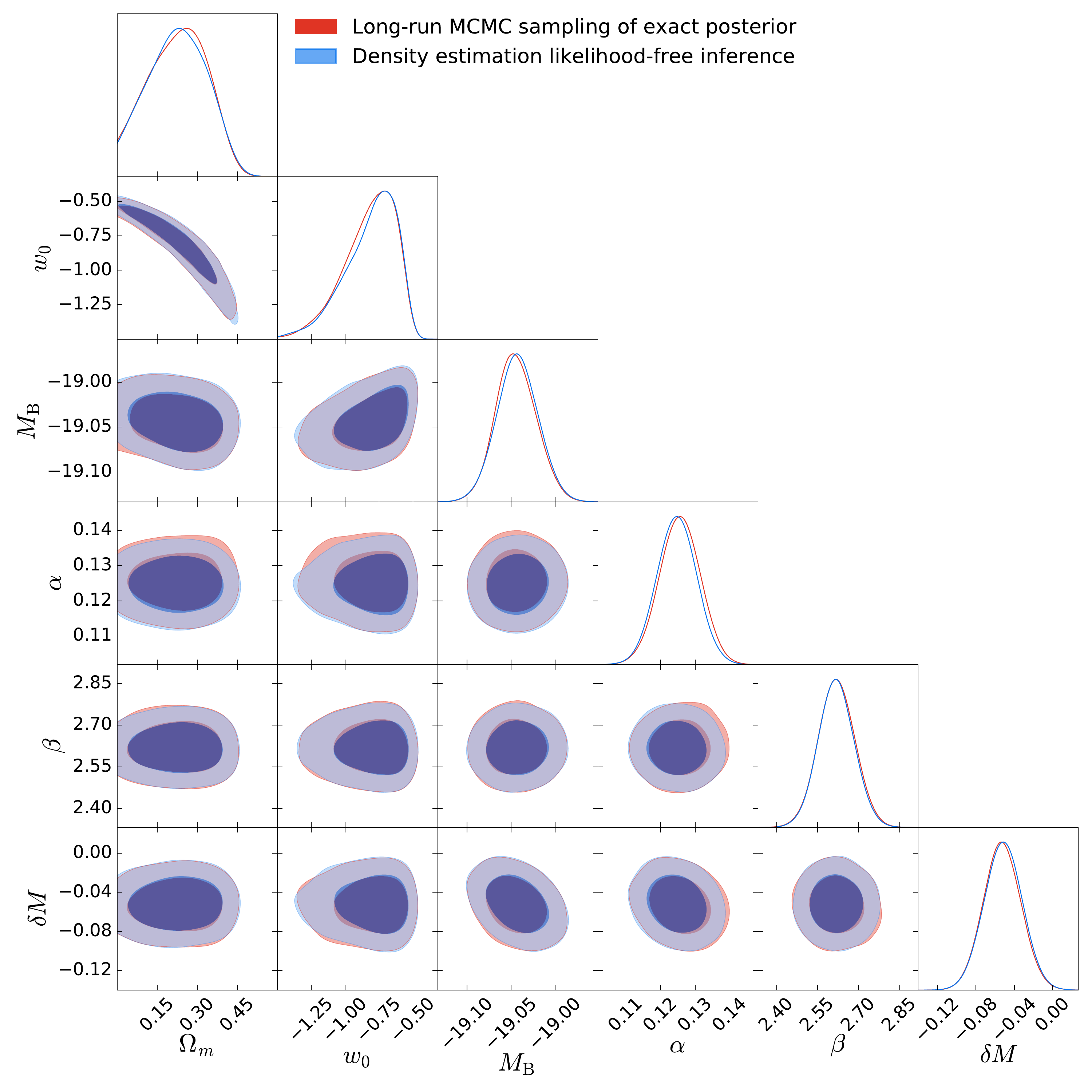}
\caption{The \textsc{delfi} posterior (blue) obtained from only $20,000$ forward simulations matches the exact posterior (red) obtained from a long-run \textsc{mcmc} chain. }
\label{fig:gmm_posterior}
\end{figure*}
\begin{figure*}
\centering
\includegraphics[width = 17.5cm]{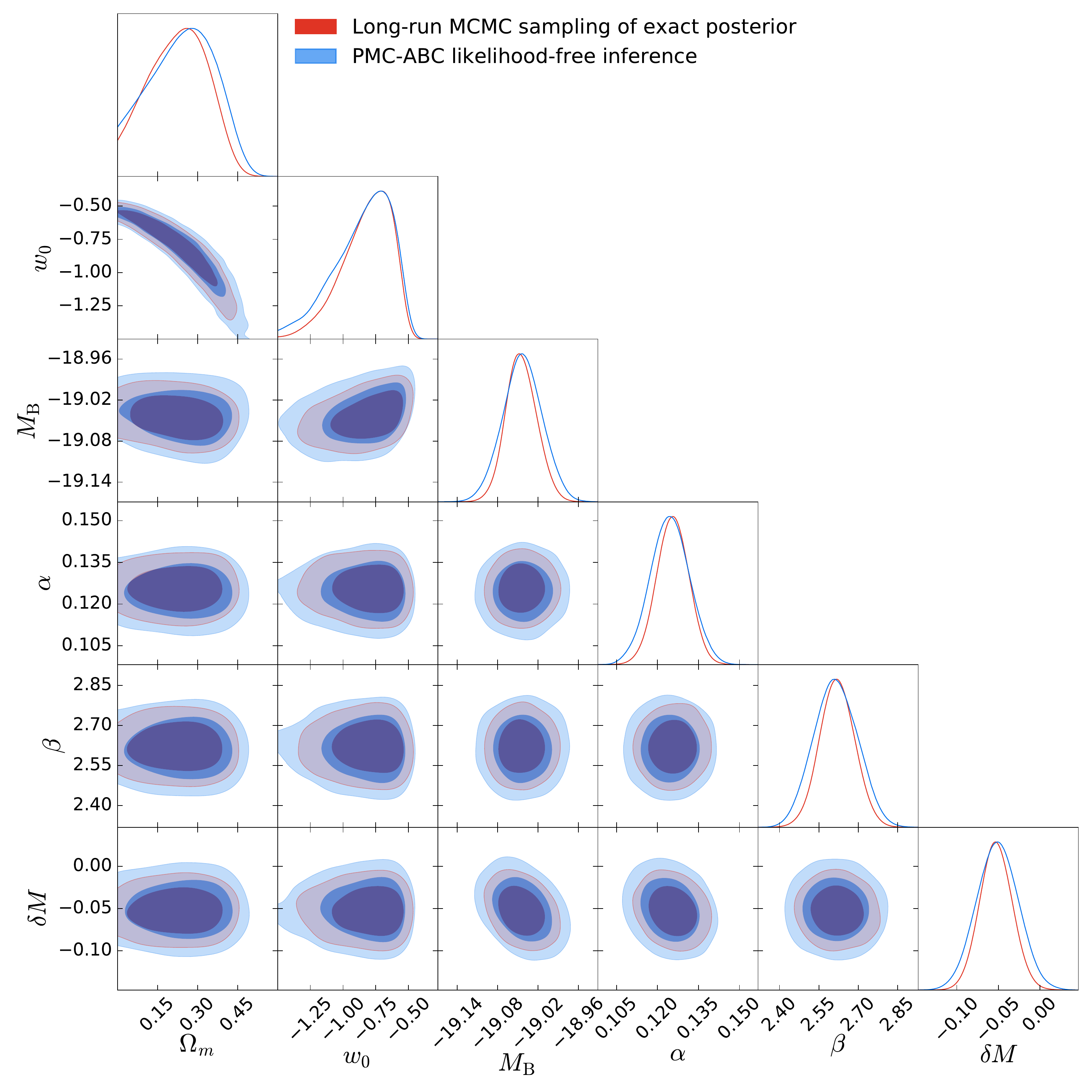}
\caption{The \textsc{pmc abc} posterior (blue) based on $20,000$ accepted samples after $> 2\times10^6$ forward simulations bounds but does not tightly approximate the exact posterior (red) obtained from a long-run \textsc{mcmc} chain.}
\label{fig:pmc_posterior}
\end{figure*}

Fig. \ref{fig:gmm_fit} shows the \textsc{gmm} fit (blue) to the samples (red). The mixture model with $12$ components is a remarkably good representation of the samples, successfully capturing the substantial and varied non-Gaussianity, degeneracies between parameters and hard prior boundaries.

Fig. \ref{fig:gmm_posterior} shows the posterior reconstruction from a long \textsc{mcmc} run on the exact posterior\footnote{$10^6$ posterior samples drawn using the affine-invariant \textsc{mcmc} code \textsc{emcee} \citep{Foreman2013}.} (red), and the posterior recovered from density-estimation likelihood-free inference (blue), following Eq. \eqref{slice} applied to the \textsc{gmm} fit shown in Fig. \ref{fig:gmm_fit}. The posterior recovered from \textsc{delfi}, using only 20,000 forward simulations, is an excellent representation of the true posterior. This is not surprising given the fidelity of the \textsc{gmm} fit to the joint data-parameter density, Fig. \ref{fig:gmm_fit}. Crucially, this also demonstrates that the massive optimal compression step -- compressing the $740$ supernovae apparent magnitudes down to just six numbers -- is for all intents and purposes, lossless; we find no perceptible increase in width of the \textsc{delfi} posteriors computed from the compressed statistics compared to the exact (\textsc{mcmc} sampled) posterior computed on the full dataset. In this validation case, since we are comparing to a ground truth with an assumed likelihood function, the compression is indeed optimal (and in this case, effectively lossless).

For comparison against the state-of-the art in \textsc{abc}, in Fig. \ref{fig:pmc_posterior} we show the recovered posterior from a long \textsc{pmc-abc} run\footnote{The \textsc{pmc-abc} implementation used here follows the algorithm described in \citet{Ishida2015}. We also tested modified \textsc{pmc-abc} algorithms following \citet{Jennings2016}, \citet{Akeret2015} and \citet{Bonassi2015}; our conclusions are unchanged by these modifications to the \textsc{pmc-abc} set-up.}. The \textsc{pmc-abc} was run through fourteen population iterations, generating $20,000$ accepted samples in the final iteration. This required $> 2\times 10^6$ forward simulations in total, since the vast majority of samples are rejected in the \textsc{pmc abc} approach. The posterior approximation obtained from the final set of samples is shown (blue) against the long run \textsc{mcmc} chain (red). The \textsc{pmc abc} run yields a reasonable approximation to the true posterior, which is unbiased but broader than the exact posterior, as expected for \textsc{abc} methods. The massive optimal data compression has enabled us to successfully perform \textsc{abc}, which would have been unfeasible in the full data-space. However, in comparison to \textsc{delfi} (Fig. \ref{fig:gmm_posterior}), \textsc{pmc abc} required $\sim 10^6$ forward simulations compared to $\sim 10^4$ for \textsc{delfi}, for a poorer approximation to the true posterior in the end.  We conclude that whilst current implementations of \textsc{abc} have limited applicability for scalable cosmological data analyses where forward simulations are expensive, \textsc{delfi} allows us to perform scalable likelihood-free Bayesian inference with a reasonable number of forward simulations (with scope for further improvement).

Finally, we estimate the Bayesian evidence using \textsc{delfi}. We separately fit a $12$ component \textsc{gmm} to the $20,000$ samples $\{\t\}$ (neglecting the $\btheta$ samples), and evaluate this estimated density at the observed data $\t_o$ to obtain the evidence. This gives an evidence estimate of $\ln\,P(\t_o) = 7.38$. In this validation case, we can compare to the evidence estimated directly from the known likelihood using nested sampling (\textsc{multinest}; \citealp{Feroz2008}). Using \textsc{multinest} we find $\ln\,P(\t_o) = 7.4(1)$, so the evidence estimates from \textsc{delfi} and \textsc{multinest} are in excellent agreement.
\section{Conclusions}
\label{sec:conclusions}
Likelihood-free inference methods allow us to perform Bayesian inference using forward simulations only, with no reference to a likelihood function. This is of particular appeal for cosmological data analysis problems where complex physical processes and instrumental effects can often be simulated, but incorporating them into a likelihood function and solving the inverse inference problem is much harder. 

Likelihood-free methods generically require large datasets to be compressed down to a small number of summary statistics in order to be scalable. We have developed a two-step compression scheme that has widespread applicability for cosmological data analysis problems. First, we compress the data down to a list of summary statistics that are carefully chosen to contain as much information about the parameters as possible, eg., compressing galaxy survey data to power spectra or other summary statistics. This type of compression is already standard practice in the analysis of cosmological surveys. Secondly, we use the optimal data compression scheme of \citet{Alsing2017} (following earlier work by \citealp{Tegmark1997} and \citealp{Heavens2000a}) to compress the list of summary statistics down to just $n$ numbers -- one per parameter -- whilst preserving the Fisher information with respect to the parameters of interest. This second compression step requires the assumption of an approximate likelihood function, and will be optimal to the extent that this is a reasonable approximation to the true (unknown) likelihood. Once the data has been compressed, all subsequent likelihood-free inference based on the massively compressed statistics is genuinely likelihood-free.

Approximate Bayesian Computation (\textsc{abc}) approaches to likelihood-free inference draw parameters from the prior and forward simulate mock data, accepting points when the simulated data fall inside some $\epsilon$-ball around the observed data. This generates samples from an approximate posterior density that becomes exact in the limit $\epsilon\rightarrow 0$. However, \textsc{abc} methods suffer from vanishingly small acceptance rates as $\epsilon\rightarrow 0$, leading to either the need for unfeasibly large numbers of forward simulations, or poor approximations to the posterior (with undesirably large values of $\epsilon$), or both.

We have presented a new approach to likelihood-free inference for cosmology -- Density-Estimation Likelihood-Free Inference (\textsc{delfi}; \citealp{Bonassi2011, Fan2013, Papamakarios2016}) -- that involves learning a parameterized model for the joint data-parameter probability density, from which (analytical models for) the posterior density and Bayesian evidence can be straightforwardly extracted. We have shown that when combined with the massive two-step data compression scheme, \textsc{delfi} is able to recover high-fidelity posterior inferences for full-scale cosmological data analyses from $\sim 10^4$ forward simulations (for a six-parameter inference task), with scope for further improvement. In contrast, current implementations of \textsc{abc} methods require orders of magnitude more forward simulations for approximate posterior inferences.

Together, massive data compression and Density Estimation Likelihood-Free Inference provide a framework for performing scalable likelihood-free inference from large cosmological datasets, even when forward simulations are computationally expensive. This opens the door to a new paradigm for principled, simulation-based Bayesian inference in cosmology.
\section*{Acknowledgements}
We thank Dan Foreman-Mackey and Boris Leistedt for useful discussions. This work is supported by the Simons Foundation. Benjamin Wandelt acknowledges support by the Labex Institut Lagrange de Paris (ILP) (reference ANR-10-LABX-63) part of the Idex SUPER, and received financial state aid managed by
the Agence Nationale de la Recherche, as part of the programme Investissements d'avenir
under the reference ANR-11-IDEX-0004-02.
	



\bibliographystyle{mnras}
\bibliography{massive}

\begin{thebibliography}{}
\makeatletter
\relax
\def\mn@urlcharsother{\let\do\@makeother \do\$\do\&\do\#\do\^\do\_\do\%\do\~}
\def\mn@doi{\begingroup\mn@urlcharsother \@ifnextchar [ {\mn@doi@}
  {\mn@doi@[]}}
\def\mn@doi@[#1]#2{\def\@tempa{#1}\ifx\@tempa\@empty \href
  {http://dx.doi.org/#2} {doi:#2}\else \href {http://dx.doi.org/#2} {#1}\fi
  \endgroup}
\def\mn@eprint#1#2{\mn@eprint@#1:#2::\@nil}
\def\mn@eprint@arXiv#1{\href {http://arxiv.org/abs/#1} {{\tt arXiv:#1}}}
\def\mn@eprint@dblp#1{\href {http://dblp.uni-trier.de/rec/bibtex/#1.xml}
  {dblp:#1}}
\def\mn@eprint@#1:#2:#3:#4\@nil{\def\@tempa {#1}\def\@tempb {#2}\def\@tempc
  {#3}\ifx \@tempc \@empty \let \@tempc \@tempb \let \@tempb \@tempa \fi \ifx
  \@tempb \@empty \def\@tempb {arXiv}\fi \@ifundefined
  {mn@eprint@\@tempb}{\@tempb:\@tempc}{\expandafter \expandafter \csname
  mn@eprint@\@tempb\endcsname \expandafter{\@tempc}}}

\bibitem[\protect\citeauthoryear{Ade et~al.,}{Ade
  et~al.}{2016}]{Planck2015XIII}
Ade P.~A.,  et~al., 2016, Astronomy \& Astrophysics, 594, A13

\bibitem[\protect\citeauthoryear{Akeret, Refregier, Amara, Seehars  \&
  Hasner}{Akeret et~al.}{2015}]{Akeret2015}
Akeret J.,  Refregier A.,  Amara A.,  Seehars S.,   Hasner C.,  2015, Journal
  of Cosmology and Astroparticle Physics, 2015, 043

\bibitem[\protect\citeauthoryear{Alsing \& Wandelt}{Alsing \&
  Wandelt}{2018}]{Alsing2017}
Alsing J.,  Wandelt B.,  2018, Monthly Notices of the Royal Astronomical
  Society: Letters, p. sly029

\bibitem[\protect\citeauthoryear{Alsing, Heavens  \& Jaffe}{Alsing
  et~al.}{2016}]{Alsing2016}
Alsing J.,  Heavens A.,   Jaffe A.~H.,  2016, Monthly Notices of the Royal
  Astronomical Society, 466, 3272

\bibitem[\protect\citeauthoryear{Arinyo-i Prats, Miralda-Escud{\'e}, Viel  \&
  Cen}{Arinyo-i Prats et~al.}{2015}]{Arinyo2015}
Arinyo-i Prats A.,  Miralda-Escud{\'e} J.,  Viel M.,   Cen R.,  2015, Journal
  of Cosmology and Astroparticle Physics, 2015, 017

\bibitem[\protect\citeauthoryear{Beaumont, Cornuet, Marin  \& Robert}{Beaumont
  et~al.}{2009}]{Beaumont2009}
Beaumont M.~A.,  Cornuet J.-M.,  Marin J.-M.,   Robert C.~P.,  2009,
  Biometrika, 96, 983

\bibitem[\protect\citeauthoryear{Betoule et~al.,}{Betoule
  et~al.}{2014}]{Betoule2014}
Betoule M. e.~a.,  et~al., 2014, Astronomy \& Astrophysics, 568, A22

\bibitem[\protect\citeauthoryear{Bishop}{Bishop}{2006}]{Bishop2006}
Bishop C.~M.,  2006, Pattern recognition and machine learning.
springer

\bibitem[\protect\citeauthoryear{Blum, Nunes, Prangle, Sisson  et~al.}{Blum
  et~al.}{2013}]{Blum2013}
Blum M.~G.,  Nunes M.~A.,  Prangle D.,  Sisson S.~A.,   et~al., 2013,
  Statistical Science, 28, 189

\bibitem[\protect\citeauthoryear{Bolton, Puchwein, Sijacki, Haehnelt, Kim,
  Meiksin, Regan  \& Viel}{Bolton et~al.}{2016}]{Bolton2016}
Bolton J.~S.,  Puchwein E.,  Sijacki D.,  Haehnelt M.~G.,  Kim T.-S.,  Meiksin
  A.,  Regan J.~A.,   Viel M.,  2016, Monthly Notices of the Royal Astronomical
  Society, 464, 897

\bibitem[\protect\citeauthoryear{Bonassi, You  \& West}{Bonassi
  et~al.}{2011}]{Bonassi2011}
Bonassi F.~V.,  You L.,   West M.,  2011, Statistical applications in genetics
  and molecular biology, 10

\bibitem[\protect\citeauthoryear{Bonassi, West  et~al.}{Bonassi
  et~al.}{2015}]{Bonassi2015}
Bonassi F.~V.,  West M.,   et~al., 2015, Bayesian Analysis, 10, 171

\bibitem[\protect\citeauthoryear{Bond, Jaffe  \& Knox}{Bond
  et~al.}{1998}]{Bond1998}
Bond J.,  Jaffe A.~H.,   Knox L.,  1998, Physical Review D, 57, 2117

\bibitem[\protect\citeauthoryear{Bond, Jaffe  \& Knox}{Bond
  et~al.}{2000}]{Bond2000}
Bond J.,  Jaffe A.~H.,   Knox L.,  2000, The Astrophysical Journal, 533, 19

\bibitem[\protect\citeauthoryear{Cameron \& Pettitt}{Cameron \&
  Pettitt}{2012}]{Cameron2012}
Cameron E.,  Pettitt A.,  2012, Monthly Notices of the Royal Astronomical
  Society, 425, 44

\bibitem[\protect\citeauthoryear{Carassou, de Lapparent, Bertin  \&
  Borgne}{Carassou et~al.}{2017}]{Carassou2017}
Carassou S.,  de Lapparent V.,  Bertin E.,   Borgne D.~L.,  2017, arXiv
  preprint arXiv:1704.05559

\bibitem[\protect\citeauthoryear{Charnock, Lavaux  \& Wandelt}{Charnock
  et~al.}{2018}]{Charnock2018}
Charnock T.,  Lavaux G.,   Wandelt B.~D.,  2018, arXiv preprint
  arXiv:1802.03537

\bibitem[\protect\citeauthoryear{Chisari et~al.,}{Chisari
  et~al.}{2018}]{Chisari2018}
Chisari N.~E.,  et~al., 2018, arXiv preprint arXiv:1801.08559

\bibitem[\protect\citeauthoryear{Davies, Hennawi, Eilers  \& Luki{\'c}}{Davies
  et~al.}{2017}]{Davies2017}
Davies F.~B.,  Hennawi J.~F.,  Eilers A.-C.,   Luki{\'c} Z.,  2017, arXiv
  preprint arXiv:1703.10174

\bibitem[\protect\citeauthoryear{Del~Moral, Doucet  \& Jasra}{Del~Moral
  et~al.}{2006}]{DelMoral2006}
Del~Moral P.,  Doucet A.,   Jasra A.,  2006, Journal of the Royal Statistical
  Society: Series B (Statistical Methodology), 68, 411

\bibitem[\protect\citeauthoryear{Fan, Nott  \& Sisson}{Fan
  et~al.}{2013}]{Fan2013}
Fan Y.,  Nott D.~J.,   Sisson S.~A.,  2013, Stat, 2, 34

\bibitem[\protect\citeauthoryear{Feeney, Mortlock  \& Dalmasso}{Feeney
  et~al.}{2017}]{Feeney2017}
Feeney S.~M.,  Mortlock D.~J.,   Dalmasso N.,  2017, arXiv preprint
  arXiv:1707.00007

\bibitem[\protect\citeauthoryear{Feroz \& Hobson}{Feroz \&
  Hobson}{2008}]{Feroz2008}
Feroz F.,  Hobson M.,  2008, Monthly Notices of the Royal Astronomical Society,
  384, 449

\bibitem[\protect\citeauthoryear{Foreman-Mackey, Hogg, Lang  \&
  Goodman}{Foreman-Mackey et~al.}{2013}]{Foreman2013}
Foreman-Mackey D.,  Hogg D.~W.,  Lang D.,   Goodman J.,  2013, Publications of
  the Astronomical Society of the Pacific, 125, 306

\bibitem[\protect\citeauthoryear{Graff, Hobson  \& Lasenby}{Graff
  et~al.}{2011}]{Graff2011}
Graff P.,  Hobson M.~P.,   Lasenby A.,  2011, Monthly Notices of the Royal
  Astronomical Society: Letters, 413, L66

\bibitem[\protect\citeauthoryear{Gualdi, Manera, Joachimi  \& Lahav}{Gualdi
  et~al.}{2017}]{Gualdi2017}
Gualdi D.,  Manera M.,  Joachimi B.,   Lahav O.,  2017, arXiv preprint
  arXiv:1709.03600

\bibitem[\protect\citeauthoryear{Gupta \& Heavens}{Gupta \&
  Heavens}{2002}]{Gupta2002}
Gupta S.,  Heavens A.~F.,  2002, Monthly Notices of the Royal Astronomical
  Society, 334, 167

\bibitem[\protect\citeauthoryear{Hahn, Vakili, Walsh, Hearin, Hogg  \&
  Campbell}{Hahn et~al.}{2017}]{Hahn2017}
Hahn C.,  Vakili M.,  Walsh K.,  Hearin A.~P.,  Hogg D.~W.,   Campbell D.,
  2017, Monthly Notices of the Royal Astronomical Society, 469, 2791

\bibitem[\protect\citeauthoryear{Heavens, Jimenez  \& Lahav}{Heavens
  et~al.}{2000}]{Heavens2000a}
Heavens A.~F.,  Jimenez R.,   Lahav O.,  2000, Monthly Notices of the Royal
  Astronomical Society, 317, 965

\bibitem[\protect\citeauthoryear{Heavens, Panter, Jimenez  \& Dunlop}{Heavens
  et~al.}{2004}]{Heavens2004}
Heavens A.,  Panter B.,  Jimenez R.,   Dunlop J.,  2004, Nature, 428, 625

\bibitem[\protect\citeauthoryear{Heavens, Sellentin, de Mijolla  \&
  Vianello}{Heavens et~al.}{2017}]{Heavens2017}
Heavens A.~F.,  Sellentin E.,  de Mijolla D.,   Vianello A.,  2017, Monthly
  Notices of the Royal Astronomical Society, 472, 4244

\bibitem[\protect\citeauthoryear{Hellwing, Schaller, Frenk, Theuns, Schaye,
  Bower  \& Crain}{Hellwing et~al.}{2016}]{Hellwing2016}
Hellwing W.~A.,  Schaller M.,  Frenk C.~S.,  Theuns T.,  Schaye J.,  Bower
  R.~G.,   Crain R.~A.,  2016, Monthly Notices of the Royal Astronomical
  Society: Letters, 461, L11

\bibitem[\protect\citeauthoryear{Hildebrandt et~al.,}{Hildebrandt
  et~al.}{2017}]{Hildebrandt2017}
Hildebrandt H.,  et~al., 2017, Monthly Notices of the Royal Astronomical
  Society

\bibitem[\protect\citeauthoryear{Hoffman, Blei, Wang  \& Paisley}{Hoffman
  et~al.}{2013}]{Hoffman2013}
Hoffman M.~D.,  Blei D.~M.,  Wang C.,   Paisley J.,  2013, The Journal of
  Machine Learning Research, 14, 1303

\bibitem[\protect\citeauthoryear{Ishida et~al.,}{Ishida
  et~al.}{2015}]{Ishida2015}
Ishida E.,  et~al., 2015, Astronomy and Computing, 13, 1

\bibitem[\protect\citeauthoryear{Jennings, Wolf  \& Sako}{Jennings
  et~al.}{2016}]{Jennings2016}
Jennings E.,  Wolf R.,   Sako M.,  2016, arXiv preprint arXiv:1611.03087

\bibitem[\protect\citeauthoryear{Joudaki et~al.,}{Joudaki
  et~al.}{2016}]{Joudaki2016}
Joudaki S.,  et~al., 2016, Monthly Notices of the Royal Astronomical Society,
  p. stw2665

\bibitem[\protect\citeauthoryear{Kacprzak, Herbel, Amara  \&
  R{\'e}fr{\'e}gier}{Kacprzak et~al.}{2017}]{Kacprzak2017}
Kacprzak T.,  Herbel J.,  Amara A.,   R{\'e}fr{\'e}gier A.,  2017, arXiv
  preprint arXiv:1707.07498

\bibitem[\protect\citeauthoryear{Kern, Liu, Parsons, Mesinger  \& Greig}{Kern
  et~al.}{2017}]{Kern2017}
Kern N.~S.,  Liu A.,  Parsons A.~R.,  Mesinger A.,   Greig B.,  2017, The
  Astrophysical Journal, 848, 23

\bibitem[\protect\citeauthoryear{Klypin, Trujillo-Gomez  \& Primack}{Klypin
  et~al.}{2011}]{Klypin2011}
Klypin A.~A.,  Trujillo-Gomez S.,   Primack J.,  2011, The Astrophysical
  Journal, 740, 102

\bibitem[\protect\citeauthoryear{Lin \& Kilbinger}{Lin \&
  Kilbinger}{2015}]{Lin2015}
Lin C.-A.,  Kilbinger M.,  2015, Astronomy \& Astrophysics, 583, A70

\bibitem[\protect\citeauthoryear{Lintusaari, Gutmann, Dutta, Kaski  \&
  Corander}{Lintusaari et~al.}{2017}]{Lintusaari2017}
Lintusaari J.,  Gutmann M.~U.,  Dutta R.,  Kaski S.,   Corander J.,  2017,
  Systematic biology, 66, e66

\bibitem[\protect\citeauthoryear{March, Trotta, Berkes, Starkman  \&
  Vaudrevange}{March et~al.}{2011}]{March2011}
March M.,  Trotta R.,  Berkes P.,  Starkman G.,   Vaudrevange P.,  2011,
  Monthly Notices of the Royal Astronomical Society, 418, 2308

\bibitem[\protect\citeauthoryear{{Marshall}, {Rajguru}  \& {Slosar}}{{Marshall}
  et~al.}{2006}]{Marshall2004}
{Marshall} P.,  {Rajguru} N.,   {Slosar} A.,  2006, \prd, 73, 067302

\bibitem[\protect\citeauthoryear{McKinley, Cook  \& Deardon}{McKinley
  et~al.}{2009}]{Mckinley2009}
McKinley T.,  Cook A.~R.,   Deardon R.,  2009, The International Journal of
  Biostatistics, 5

\bibitem[\protect\citeauthoryear{Melchior \& Goulding}{Melchior \&
  Goulding}{2016}]{Melchior2016}
Melchior P.,  Goulding A.~D.,  2016, arXiv preprint arXiv:1611.05806

\bibitem[\protect\citeauthoryear{Mesinger, Greig  \& Sobacchi}{Mesinger
  et~al.}{2016}]{Mesinger2016}
Mesinger A.,  Greig B.,   Sobacchi E.,  2016, Monthly Notices of the Royal
  Astronomical Society, 459, 2342

\bibitem[\protect\citeauthoryear{Panter, Jimenez, Heavens  \& Charlot}{Panter
  et~al.}{2007}]{Panter2007}
Panter B.,  Jimenez R.,  Heavens A.~F.,   Charlot S.,  2007, Monthly Notices of
  the Royal Astronomical Society, 378, 1550

\bibitem[\protect\citeauthoryear{Papamakarios \& Murray}{Papamakarios \&
  Murray}{2016}]{Papamakarios2016}
Papamakarios G.,  Murray I.,  2016, in Advances in Neural Information
  Processing Systems. pp 1028--1036

\bibitem[\protect\citeauthoryear{{Protopapas}, {Jimenez}  \&
  {Alcock}}{{Protopapas} et~al.}{2005}]{Protopapas2005}
{Protopapas} P.,  {Jimenez} R.,   {Alcock} C.,  2005, \mnras, 362, 460

\bibitem[\protect\citeauthoryear{Reichardt, Jimenez  \& Heavens}{Reichardt
  et~al.}{2001}]{Reichardt2001}
Reichardt C.,  Jimenez R.,   Heavens A.~F.,  2001, Monthly Notices of the Royal
  Astronomical Society, 327, 849

\bibitem[\protect\citeauthoryear{Riess et~al.,}{Riess et~al.}{2011}]{Riess2011}
Riess A.~G.,  et~al., 2011, The Astrophysical Journal, 730, 119

\bibitem[\protect\citeauthoryear{Robin, Reyl{\'e}, Fliri, Czekaj, Robert  \&
  Martins}{Robin et~al.}{2014}]{Robin2014}
Robin A.,  Reyl{\'e} C.,  Fliri J.,  Czekaj M.,  Robert C.,   Martins A.,
  2014, Astronomy \& Astrophysics, 569, A13

\bibitem[\protect\citeauthoryear{Rubin et~al.}{Rubin et~al.}{1984}]{Rubin1984}
Rubin D.~B.,  et~al., 1984, The Annals of Statistics, 12, 1151

\bibitem[\protect\citeauthoryear{Schafer \& Freeman}{Schafer \&
  Freeman}{2012}]{Schafer2012}
Schafer C.~M.,  Freeman P.~E.,  2012, in , Statistical Challenges in Modern
  Astronomy V.
Springer, pp 3--19

\bibitem[\protect\citeauthoryear{Sellentin \& Heavens}{Sellentin \&
  Heavens}{2015}]{Sellentin2015}
Sellentin E.,  Heavens A.~F.,  2015, Monthly Notices of the Royal Astronomical
  Society: Letters, 456, L132

\bibitem[\protect\citeauthoryear{Sisson, Fan  \& Tanaka}{Sisson
  et~al.}{2007}]{Sisson2007}
Sisson S.~A.,  Fan Y.,   Tanaka M.~M.,  2007, Proceedings of the National
  Academy of Sciences, 104, 1760

\bibitem[\protect\citeauthoryear{Springel}{Springel}{2005}]{Springel2005}
Springel V.,  2005, Monthly notices of the royal astronomical society, 364,
  1105

\bibitem[\protect\citeauthoryear{Springel et~al.,}{Springel
  et~al.}{2017}]{Springel2017}
Springel V.,  et~al., 2017, Monthly Notices of the Royal Astronomical Society

\bibitem[\protect\citeauthoryear{Tegmark, Taylor  \& Heavens}{Tegmark
  et~al.}{1997}]{Tegmark1997}
Tegmark M.,  Taylor A.~N.,   Heavens A.~F.,  1997, The Astrophysical Journal,
  480, 22

\bibitem[\protect\citeauthoryear{Toni \& Stumpf}{Toni \&
  Stumpf}{2009}]{ToniStumpf2009}
Toni T.,  Stumpf M.~P.,  2009, Bioinformatics, 26, 104

\bibitem[\protect\citeauthoryear{Toni, Welch, Strelkowa, Ipsen  \& Stumpf}{Toni
  et~al.}{2009}]{Toni2009}
Toni T.,  Welch D.,  Strelkowa N.,  Ipsen A.,   Stumpf M.~P.,  2009, Journal of
  the Royal Society Interface, 6, 187

\bibitem[\protect\citeauthoryear{Tripp}{Tripp}{1998}]{Tripp1998}
Tripp R.,  1998, Astronomy and Astrophysics, 331, 815

\bibitem[\protect\citeauthoryear{Weyant, Schafer  \& Wood-Vasey}{Weyant
  et~al.}{2013}]{Weyant2013}
Weyant A.,  Schafer C.,   Wood-Vasey W.~M.,  2013, The Astrophysical Journal,
  764, 116

\bibitem[\protect\citeauthoryear{Zablocki \& Dodelson}{Zablocki \&
  Dodelson}{2016}]{Zablocki2016}
Zablocki A.,  Dodelson S.,  2016, Physical Review D, 93, 083525

\makeatother
\end{thebibliography}

\bsp	
\label{lastpage}
\end{document}